\documentclass[aps,prb,twocolumn,superscriptaddress,showpacs]{revtex4}
\usepackage{epsfig}
\usepackage[english]{babel}
\usepackage{graphics}
\usepackage{amssymb}
\usepackage{amsmath}

\begin{document}

\title{Magnetoresistance Anisotropy of Polycrystalline Cobalt Films:\\ Geometrical-Size- and Domain-Effects}

\author{Woosik Gil}
\affiliation{Institut f\"ur Angewandte Physik und Zentrum f\"ur Mikrostrukturforschung, Universit\"at Hamburg, Jungiusstrasse 11, D-20355 Hamburg, Germany}
\author{Detlef G\"orlitz}
\affiliation{Institut f\"ur Angewandte Physik und Zentrum f\"ur Mikrostrukturforschung, Universit\"at Hamburg, Jungiusstrasse 11, D-20355 Hamburg, Germany}
\author{Michael Horisberger}
\affiliation{ETH Z\"urich \& Paul Scherrer Institut, CH-5232 Villigen PSI, Switzerland}
\author{J\"urgen K\"otzler}
\affiliation{Institut f\"ur Angewandte Physik und Zentrum f\"ur Mikrostrukturforschung, Universit\"at Hamburg, Jungiusstrasse 11, D-20355 Hamburg, Germany}


\date{\today}

\begin{abstract}
The magnetoresistance (MR) of 10~nm to 200~nm thin polycrystalline Co-films, deposited on glass and insulating Si(100), is studied in fields up to 120~kOe, aligned along the three principal directions with respect to the current: longitudinal, transverse (in-plane), and polar (out-of-plane). At technical saturation, the anisotropic MR (AMR) in polar fields turns out to be up to twice as large as in transverse fields, which resembles the yet unexplained geometrical size-effect (GSE), previously reported for Ni- and Permalloy films. Upon increasing temperature, the polar and transverse AMR's are reduced by phonon-mediated sd-scattering, but their ratio, i.e. the GSE remains unchanged. Basing on Potters's theory [Phys.Rev.B \textbf{10}, 4626(1974)], we associate the GSE  with an anisotropic effect of the spin-orbit interaction on the sd-scattering of the minority spins due to a film texture. Below magnetic saturation, the magnitudes and signs of all three MR's depend significantly on the domain structures depicted by magnetic force microscopy. Based on hysteresis loops and taking into account the GSE within an effective medium approach, the three MR's are explained by the different magnetization processes in the domain states. These reveal the importance of in-plane uniaxial anisotropy and out-of-plane texture for the thinnest and thickest films, respectively.
\end{abstract}
\pacs{73.50.-h; 73.50.Jt; 73.61.-r; 75.47.-m}


\maketitle

\section{Introduction}
In applied magnetism, the coupling of the magnetic moment to spatial degrees of freedom plays a key role, and this especially applies to modern magneto- or spin-electronics\cite{TRM75}. Basically, this coupling is provided by the spin-orbit interaction, which in the example of magnetotransport causes the scattering rate of the conduction electrons, $\tau^{-1}$, to depend on the direction of the local magnetization $\vec{M}$ with respect to the current. In the archetypal bulk ferromagnets iron, cobalt, nickel, and their alloys, the resistance difference for orientations of $\vec{M}$ parallel and perpendicular to the current, i.e. the socalled anisotropic magnetoresistance ratio (AMR), $\Delta\rho/\rho\equiv\left(\rho_\parallel~-~\rho_{\bot}\right)/\rho$, amounts to some percent. In nanostructured devices like magnetic multilayers, wires, or constrictions in the ballistic regime\cite{VSJ05}, this ratio may be enhanced to several ten percent. 

Basically, the determination of the scattering rate $\tau^{-1}(\vec{M})$ and of the AMR requires the knowledge of the scattering potential and also of the spin-orbit split bands at the Fermi-surface $\in_F$. Some special aspects of the AMR have been evaluated by Smit\cite{Smi51}, Berger\cite{Ber64}, Potter\cite{Pot74}, and Fert and Campbell\cite{FC76}, however, the evaluation of $\tau^{-1}(\vec{M})$ for a realistic case is still lacking, at least to the best of our knowledge. In this context, we note a recent \textit{ab initio} calculation of the intrinsic anomalous Hall-effect which, in contrast to the AMR, depends only to first order on the spin-orbit interaction and \emph{not} on a scattering potential. This quantity was obtained by integrating the k-space Berry-phase over the occupied spin-orbit split states of iron\cite{YKM04} and was found to agree up to some 30 percent with data on iron whiskers\cite{Dhe67}.

The present work is intended to a fairly systematic study of the AMR, which is of second-order in the spin-orbit interaction, also in an elemental 3d-ferromagnet. By selecting hcp cobalt with a rather well known band-structure\cite{Pap86}, some deeper insight into the AMR may be facilitated. By choosing polycrystalline films, we are closer to devices which invariably use polycrystalline materials. We will vary the structural disorder and the temperature in the films to probe the role of different scattering mechnisms. These basic properties of the films under study are examined in Sect.II. Section III is devoted to the AMR in the technically saturated state with main emphasis to a still unexplained phenomenon of the AMR, i.e. the socalled geometrical-size effect (GSE), previously observed in thin Ni\cite{CM72} and Permalloy\cite{RLC97} films. Another point of interest will be the absolute value of $\Delta\rho$ at low temperatures: for Ni-alloys, already McGuire and Potter\cite{MP75} pointed out the unsensitivity of $\Delta\rho$ against significant variations of the residual resistivity $\rho(0)$. The influence of different domain structures, depending on the film thicknesses, on the magnetoresistance, is investigated in Section IV and will be discussed by using the results on the GSE. This low-field regime, where the in-plane AMR switches at rather small coercive fields $(H_c\approx10~\mbox{Oe})$, may be of interest for applications despite the fact that $\Delta\rho/\rho$ lies in the range of some percent. The summary and conclusions are contained in Section V.

\section{Characterization of the films}

\begin{figure}[t]
\psfig{file=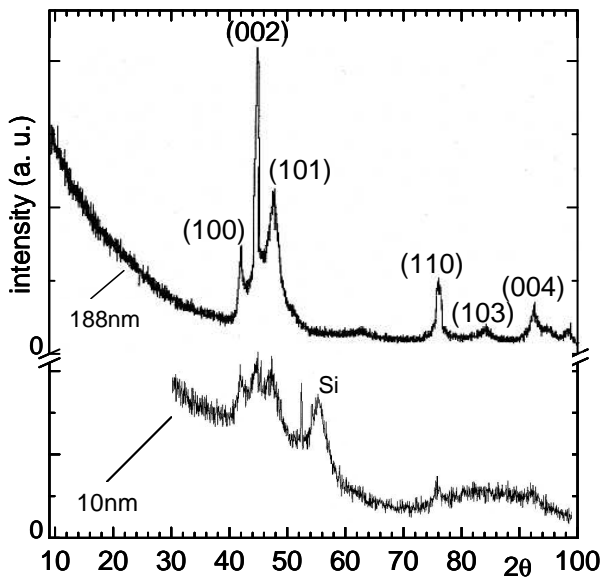, width=8cm}
\caption{\protect{Fig.1. Wide-angle X-ray diffraction patterns of Co-films deposited by DC-magnetron sputtering on glass (188~nm) and on oxidized Si (100) (10~nm). The Miller indices ($hk\ell$) denote the reflections expected for the hcp structure at incident wavelength $\lambda_{CuK_{\alpha}} = 1.54$~\AA .}}
\label{Fig1}
\end{figure}

By means of DC-magnetron sputtering at an Ar-pressure of 2$\cdot$10$^{-9}$~bar, cobalt films of thicknesses 10~nm, 20~nm, and 188~nm were deposited on Synsil-glass and oxidized Si(100) surfaces and capped by 3~nm thick Al-layers. The thicknesses were measured by a profilometer to an accuracy of 0.6~nm and controlled by high-resolution SQUID-magnetometry. X-ray diffraction diagrams (XRD), as shown in Fig.1, revealed a polycrystalline hcp-structure with a slight texture of the hexagonal axis normal to the plane. Surface images recorded by an atomic force microscope (AFM, Q-Scope$^{\textup{TM}}$250, Quesant Instruments Co.) yielded surface roughnesses between $(1.5\pm0.3)$~nm for 10~nm and $(3.8\pm0.5)$~nm for 188~nm and indicated the grain sizes to increase from $(25\pm5)$~nm to $(80\pm5)$~nm. Within the error margins, these results turned out to be the same for both substrates. It is interesting to note that the grain sizes and their increase with thickness are consistent with a recent report for polycrystalline Co on glass and Si(100) substrates \cite{KCB04}.

The magnetic properties of all films have been investigated by ferromagnetic resonance (FMR), hysteresis loops, and magnetic force microscopy (MFM). Using a home-made FMR spectrometer operating at 9.1 GHz, the directions and magnitudes of small uniaxial anisotropy fields, $\vec{H}_u$, in the film planes were determined . On 20~nm Co:Si and 188~nm Co:glass, for example,  $H_u=22.3$~Oe and 15.3~Oe, respectively, was obtained and the orientation of $\vec{H}_u$ could be related to the direction of the deposition process. Magnetization isotherms were measured by a SQUID-magnetometer (Quantum Design MPMS2) along three orthogonal directions of the applied field $\vec{H}$ at temperatures, which were of interest for the analyses of the magnetoresistances (MR's) in Section IV. There also MFM images are presented in order to visualize the domain structure underlying the magnetization processes, see Fig.7 below. For this purpose, the Q-scope was equipped with a commercial tip, coated by a 40~nm thin hard Co-alloy (Nanosensors$^{ \textup{TM}}$), and magnetized perpendicularly to the scanning directions. The directions of the in-plane magnetization were determined by monitoring the domain wall motion induced by a small magnetic field produced by external Helmholtz coils.

The resistances have been measured by an array of four in-line contacts prepared parallel to $\vec{H}_u$ by ultrasonic bonding. The driving currents were kept small enough to produce linear responses and the resulting $U/I$-ratios were corrected for the sample geometry \cite{Smi58} to determine the resistivities of the films. The sample chip was mounted to the end of an cold-finger extending from the cold-plate of a pulse-tube cooler (PRK, Transmit Co. Giessen, Germany) to the center of a warm-bore superconducting magnet (130~kOe, Oxford Instruments). A PID controller and a heater allowed stable sample temperatures between 70~K and 350~K. Measurements of the magnetoresistance in the domain states, i.e. at low magnetic fields, were performed by means of an electromagnet, by which also the angle between current and field could be varied. More experimental details are given in Ref.\onlinecite{Gil04}. We should mention here, that the structural, magnetic, and transport properties proved to be largely independent on the substrate, i.e. synsil-glass or oxidized Si(100)\cite{Gil04}. This feature indicates a dominant effect of the polycrystallinity of the films, i.e. of the deposition process. For some practical reasons, we selected three films with thicknesses between 10~nm and 188~nm for the present study.

The temperature dependence of the zero-field resistivities is depicted in Fig.2 for these three films. The data can be well described by a sum of three contributions

\begin{equation}
\rho(T)=\rho(0)+\rho_{ph}(T)+\rho_{m}(T)  .
\end{equation}
\noindent
According to the inset, the residual resistivity increases linearly with the inverse thickness, 

\begin{equation}\nonumber
\rho(0)=\rho_b(0)[1+d_c/d].
\end{equation}
\noindent
The characteristic thickness, $d_c=(18\pm1)~\mbox{nm}$ cannot be related to an extra scattering by the film surfaces\cite{Son52} or grain boundaries\cite{MS70} since, the theories predict the 1/d-behavior only for small deviations from the bulk limit, $\rho(0)\geq\rho_b(0)$. Hence, the observed increase of $\rho(0)$ indicates scattering by an additional, yet unidentified disorder in the thinner films. Using the extrapolated bulk value, $\rho_b(0)=(11+1)~\mu\Omega\mbox{cm}$, the carrier density $5.8\cdot10^{22}\mbox{cm}^{-3}$ from Hall-data for these films\cite{Gil04}, and the free electron model for the conduction electrons in Co\cite{GSN93}, we find an upper limit for the mean free path, $\ell_e \left(0 \right) = \hbar k_F/n_e e^{2}\rho_b \left(0 \right)\approx 11~\mbox{nm}$. Since this length is significantly smaller than the mean grain sizes observed by AFM, it may be associated with point-defect scattering within the otherwise crystalline grains.

Since the electron-magnon scattering in Co, $\rho_m(T)=1.5\cdot10^{-5}(\mu\Omega\mbox{cm~K}^{-2})~T^{2}\quad$ \cite{RVS02} is small, the temperature variation of the resistivities should be dominated by phonons. Indeed, by fitting $\rho(T)$ to the Bloch-Grueneisen form

\begin{equation}\nonumber
\rho_{ph}(T)=\rho_{ph}\cdot\left({\frac{T}{{\Theta _D}}}\right)^n \,\,\int\limits_0^{\Theta_D/2T}{\frac{{x^{n}}}{{\left( {\sinh \,x}\right)^{2}}}} \,dx
\end{equation} 
\noindent
and taking for the Debye temperature $\Theta_D$=445~K, we find an excellent agreement by setting for the exponent n=3, valid for phonon-mediated sd-scattering\cite{Wil38,Goo63}. The strength of this scattering, $\rho_{ph}$, becomes smaller in the thinner, more disordered films, however, due to coupling of the phonons to the complicated structure of the d-states, it is difficult to estimate, $\rho_{ph}$, even for single crystals\cite{Goo63}.

Finally, it may be interesting to note that the present resistivities of the 188~nm film are almost identical to those obtained by Freitas \textit{et al.} \cite{FGM90} on a 300~nm Co film deposited by magnetron sputtering on glass. This applies to the residual resistivity, $\rho(0)=14~\mu\Omega\mbox{cm}$, as well as to $\rho(T)$ at room temperature. Significantly  smaller $\rho(0)$-values have been detected on diode sputtered\cite{FGM90} and epitaxial\cite{RYT99} films of similar thickness.
\begin{figure}[t]
\psfig{file=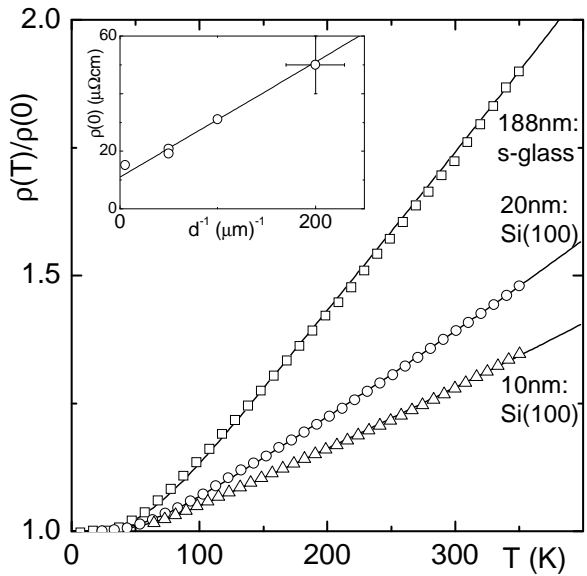, width=8cm} 
\caption{\protect{Temperature variation of the zero-field resistance of the three Co-films under study. The solid line represents a fit to Equ.1, taking into account the contributions by phonon-mediated sd- and electron-magnon scattering. Inset: linear dependence of the residual resistivity on the inverse thickness d, including a result for d=5~nm from Ref.\onlinecite{Pot74}.}}
\label{Fig2}
\end{figure}

\section{High-Field Magnetoresistance}

\begin{figure}[t]
\psfig{file=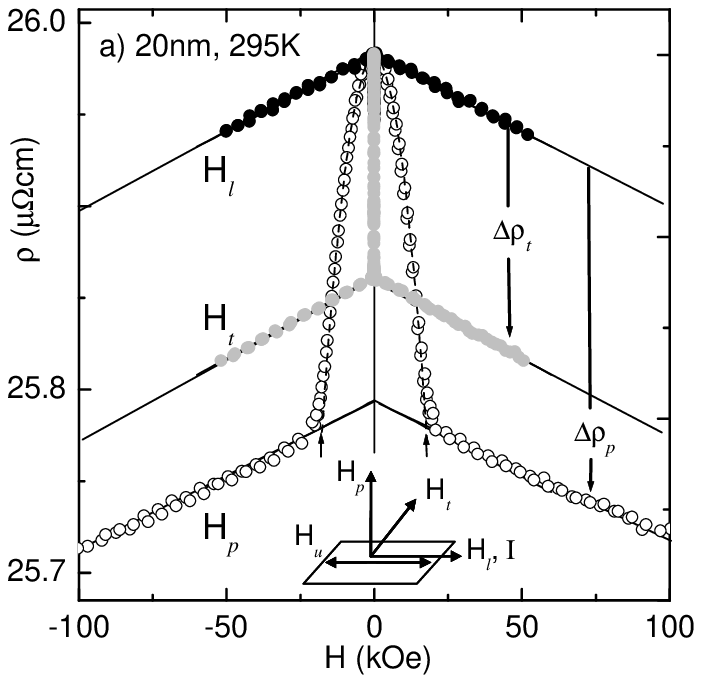, width=7.1cm}
\psfig{file=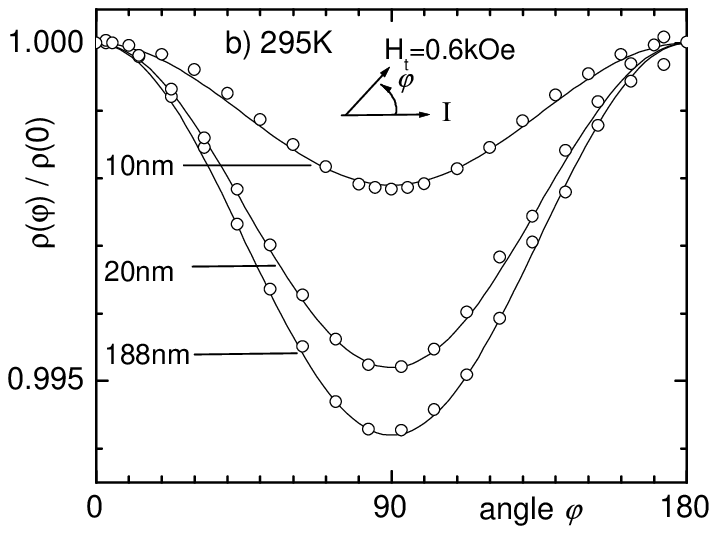, width=7.1cm}
\caption{\protect{a) High-field magnetoresistance (MR) of the 20~nm Co-film at room temperature for the three principal directions of the field (see inset), revealing  the transverse and polar MR's, $\Delta \rho_t$ and $\Delta \rho_p$, and the appearance of a linear negative MR at $M>M_s$. The dashed curve through the polar MR-data presents a fit to Equ.2. b) Normalized in-plane MR of Co-films, $\rho(\varphi)/\rho(0)$, recorded at room temperature as a function of the angle $\varphi$ between current and field above the saturation field of the in-plane magnetization. The solid curves represent fits to Equ.2 valid for the anisotropic MR (AMR).}}
\label{Fig3}
\end{figure}

The MR of all films has been studied for three principal directions of $\vec{H}$, defined by the directions of the current $(\vec{I}||\vec{H}_u)$ and the film plane, see inset to Fig.3a. To give an example, Fig.3a shows the three MR's of the 20~nm film at room-temperature. Starting from a common value at low fields, a negative MR is found for all directions of $\vec{H}$. While the longitudinal MR, $\rho(H_{\ell})$ decreases linearly with field, the transverse and polar MR's contain additional contributions. Above the saturation fields $H_s$, where the films become homogeneously magnetized, $\vec{M}(H>H_s)=M_s\vec{H}/H$, these additional contributions to the MR also saturate at values $\Delta\rho_t=\rho_{\ell}-\rho_t$ and $\Delta\rho_p=\rho_{\ell}-\rho_p$, both indicated in Fig.3a. This contribution results from the spin-orbit induced AMR, since upon rotation of the magnetization either to the transverse or to the polar direction, we realize the angular dependence

\setcounter{equation}{1}
\begin{equation}
\rho(\varphi)=\rho(0)-\Delta\rho\cdot sin^{2}\varphi ,
\end{equation}

\noindent
where $\varphi$ is the angle between current and the direction of the magnetization $\vec{M}$. Such behavior is characteristic of the AMR of polycrystalline samples of cubic or hexagonal ferromagnets\cite{MP75}, and is illustrated by Fig.3b for the in-plane rotation of $\vec{M}$ in a field $H=0.6~\mbox{kOe}>H_s$. Details of the MR during saturation by (weak) \textit{in-plane} fields will be discussed in Section IV. Here we look at the \textit{polar} MR by increasing $H_p$ in Fig.3a. SQUID magnetization data\cite{Gil04} reveal $M_p(H_p < M_s)=H_p$ due to a rotation of $\vec{M}$ against the in-plane demagnetizing field $H_N=-M_s$, so that the angle between $\vec{M}$ and current $\vec{I}(||\vec{H}_u)$ is determined by $sin~\varphi=M_p/M_s=H_p/M_s$. For this case, Equ.2 predicts a parabolic decrease, $\rho(H_p)-\rho(0)=-\Delta\rho_p(H_p/M_s)^{2}$, which is depicted in Fig.3a by the dashed curve in full agreement with the data.

\subsection{Spin-wave contribution}

\begin{figure}[t]
\center
\psfig{file=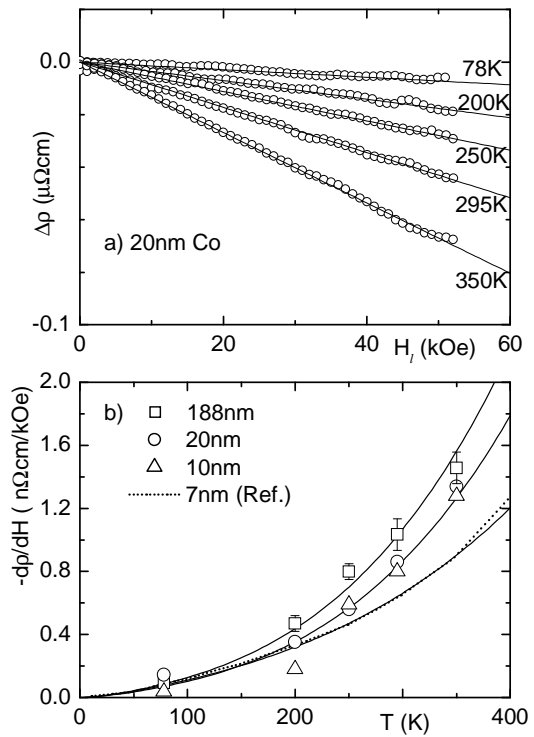, width=7.1cm}
\caption{\protect{a) Longitudinal high-field MR of the 20~nm Co-film at various temperatures between 78~K and 350~K. b) Coefficient of the linear high-field MR (see panel a.) for all films \textit{vs.} temperature. For comparision, the dotted curve shows results for a 7~nm polycrystalline Co-film from Ref.\onlinecite{RVS02}, while the solid curves are fits to the prediction for scattering by spin-waves, Equ.4.}}
\label{Fig4}
\end{figure}
It is evident from Fig.3a that the linear MR, $d\rho/dH$, is the same in all directions of $\vec{H}$. No signature of the classical Lorentz-MR, which is positive and proportional to $(M+H)^{2}$, is realized for $H>H_s$, even not at room temperature. Due to the small mean free path the absence of this effect is plausible, while in epitaxial films it becomes visible\cite{RYT99}. The linear MR has been realized before in $\rho(H_{\ell})$ on epitaxially grown iron, cobalt, and nickel films on $MgO$ and $Al_2O_3$\cite{RVS02} and was quantitatively discussed in terms of elastic scattering by thermally excited magnons. Roughly spoken, the negative MR can be ascribed to the suppression of low energy magnons, which results from the increase of the magnon gap proportional to H. The strong thermal increase of $d\rho/dH$ is illustrated by Fig.4a for the longitudinal MR to which the AMR does not contribute. In Fig.4b, their temperature dependence is shown for the three films under study and compared to the result for a 7~nm thin Co-film obtained by Raquet \textit{et al.}\cite{RVS02}. These authors fitted their data to a simplified model for sd-scattering by magnons\cite{Goo63},

\begin{equation}
\frac{{d\rho }}
{{dH}} = A~T\left( {1 + 2d_1 T^2 } \right)\, \ln \left( {\frac{T}
{{T_0 }}} \right), 
\end{equation}

\noindent
where the amplitude A changed only little from 3 to 4 $p\Omega\mbox{cm/K~kOe}$. Since A depends on the sd-exchange, numerical estimates are rather difficult. The coefficient $d_1=D_1/D_0$ is determined by the ratio of the mass renormalization coefficient $D_1$ and the zero-temperature stiffness of the spin-waves $D_0$. Independent experimental data for Co yield $d_1=1.57\cdot 10^{-6}~\mbox{K}^{-2}$ in good agreement with calculations, and it was argued\cite{RVS02} that $d_1$ might be rather insensitive to microstructural details of the films. Consequently, we fitted our data to Equ.3 admitting (plausible) variations in the amplitude A and found a larger value, $d_1=3\cdot10^{-6}\mbox{K}^{-2}$. We believe that the difference is related to the rather strong disorder in the present films with a residual resistance ratio (RRR) near 2 (see Fig.2), which contrasts to RRR=27 reported by Raquet \textit{et al.}\cite{RVS02} for their thickest films. Hence, one may suspect that the granular disorder in our films gives rise to a stronger thermal renormalization of 'the spin-wave energies'.

\subsection{Anisotropic Magnetoresistance}

\begin{figure}[ht]
\center
\psfig{file=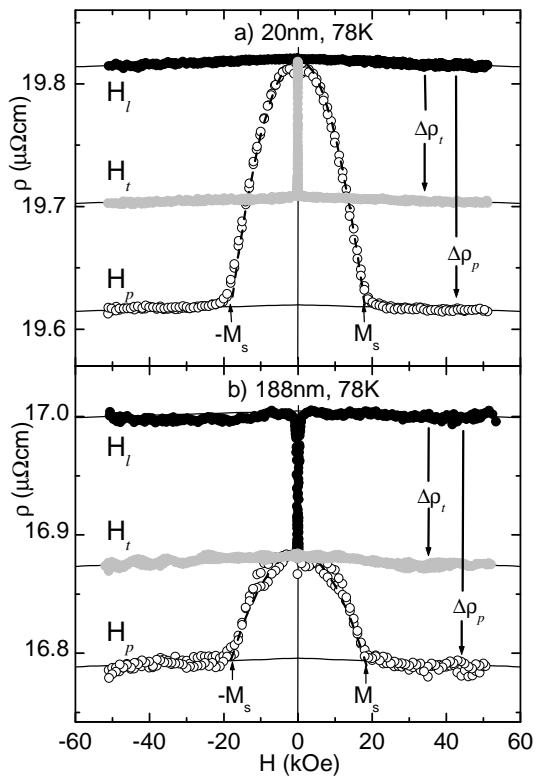, width=7.1cm}
\caption{\protect{High-field MR of a) 20~nm Co- and b) 188~nm Co-films, measured at 78~K for the three principal directions of ${\vec{H}}$ relative to the current. As for the in-plane MR, see Fig.3b, the quadratic decrease of the polar MR's, $\rho_p$($H$), observed for $M<M_s$ (dot curves) signalizes the AMR-effect also. Above technical saturation, $M(H,T)\geq M_s$, the saturation values of the transverse $(\Delta \rho_t)$ and the polar $(\Delta \rho_p)$ AMR's are indicated.}}
\label{Fig5}
\end{figure}
At low temperatures, where the spin-wave contribution vanishes, the AMR effect should prevail. This is demonstrated in Fig.5 by the MR curves of the 20~nm and 188~nm films measured at T=78~K along the three principal directions of the field. The significant difference between the MR's of both films at smaller fields is related to the domain structure and will be discussed in Section IV. Here we focus on the saturated transverse and polar AMR's, $\Delta\rho_t$ and $\Delta\rho_p$, which differ significantly from each other, but do not change very much with thickness (essentially the same observation is made on the 10~nm film). This phenomenon is one of our main results: for all thicknesses, the polar AMR turns out to be about twice as large as the transverse AMR.

At first, a sizable difference between both MR's, $\Delta\rho_p>\Delta\rho_t$, has been reported by Chen and Marsocci\cite{CM72} for single- and poly-crystalline nickel films. They coined this feature as 'geometrical size effect' (GSE) and believed that it may arise from the electronic structure inside the film material. More recently, this size-effect has also been detected on sputter-deposited 4.5~nm to 100~nm thin Permalloy films\cite{RLC97} at a low temperature, T=5~K. This study revealed that by raising the degree of (111)-texture in the film, $\Delta\rho_p$ was increased so that the ratio $\Delta\rho_p/\Delta\rho_t$ tended towards two. An attempt to explain this GSE by assuming an anisotropic scattering rate due to diffuse scattering at the film boundaries, however, did not provide conclusive results\cite{RLC97}.

\begin{figure}[ht]
\center
\psfig{file=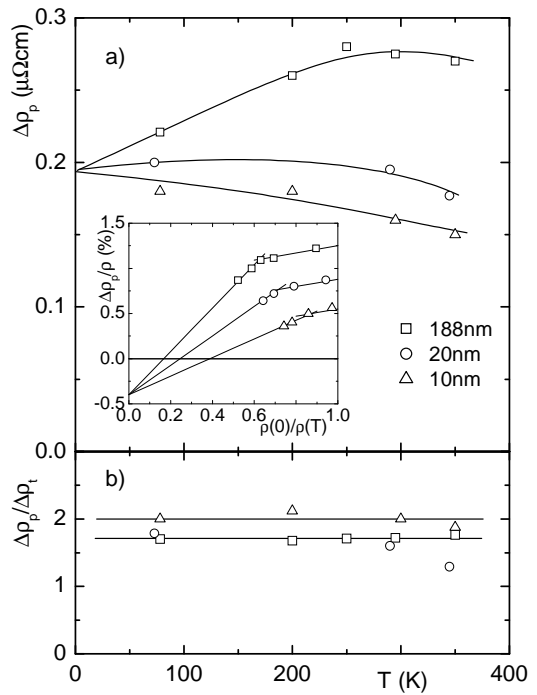, width=7.1cm}
\caption{\protect{Temperature dependences of a) the saturation values of the polar magnetoresistivity $\Delta\rho_p$ and b) of the geometrical size-effect. All solid curves are guides to the eye. Inset to a): Parker-plot analysis of $\Delta\rho_p/\rho$; the straight lines through the data indicate a common negative phonon-contribution to the MR.}}
\label{Fig6}
\end{figure}
In order to explore the AMR and the GSE of our Co-films in some more detail, the absolute values and the thermal behavior of both $\Delta\rho's$ are summarized in Fig.6. Two remarkable features should be emphasized: (i) despite different temperature variations, the MR's of all films can be extrapolated to the same value at T=0, as shown in Fig.6a for the polar direction; (ii)  Fig.6b demonstrates that the GSE, i.e. the ratio of the polar and transverse AMR's, remains almost independent of temperature. 

At first, we address to the AMR postponing the discussion of the GSE to the following subsection. A thickness-independence of $\Delta\rho$ itself rather than of the ratio $\Delta\rho / \rho$ has been pointed out earlier for $Ni_{0.7}Co_{0.3}$ and $Ni_{0.8}Fe_{0.2}$ alloys (see Fig.17 of Ref.\onlinecite{MP75}). For all present Co-films, $\Delta\rho_p(0)=0.19~\mu\Omega\mbox{cm}$ follows from Fig.6a, and we suspect that the origin of this AMR resides in the crystalline regions, to which we tentatively assigned already the bulk residual resistivity, $\rho_b(0)=11~\mu\Omega\mbox{cm}$, in Sect.II. There we determined the mean-free path, $\ell_e=11~\mbox{nm}$, which turned out to be much smaller than the grain sizes estimated from AFM images\cite{Gil04}. Therefore, we relate the low-temperature AMR $\Delta\rho(T \to 0)$ also to the scattering within the crystalline grains and believe that the extra scattering, which enhances $\rho(0)$ in the thinner films (see inset to Fig.2), produces a negligible AMR. In fact, a weak AMR is expected for scattering potentials with reduced symmetry, e.g. associated with phonons\cite{MP75} or correlated structural disorder (grain boundaries, dislocations), because in these cases the directional symmetry-breaking effect by the magnetization $\vec{M}$ via the spin-orbit interaction becomes less effective.

Analyzing the effect of temperature, i.e. of phonon-scattering, we employ the widely used Parker-plot \cite{Par51}, based on the relation for the AMR ratio,

\begin{equation}
\frac{{\Delta \rho \left( T \right)}}
{{\rho \left( T \right)}} = \left[ {\left( {\frac{{\Delta \rho }}
{\rho }} \right)_d  - \left( {\frac{{\Delta \rho }}
{\rho }} \right)_{T} } \right]\,\,\,\frac{{\rho \left( 0 \right)}}
{{\rho \left( T \right)}} + \left( {\frac{{\Delta \rho }}
{\rho }} \right)_T  .
\end{equation}
\\

\noindent
This equation is valid under the two premises: (i) the electric transport is dominated by one spin-channel, i.e. the majority channel in Co\,\,\cite{GSN93}, and (ii) Matthiessen's rule applies for the thermal and defect scattering. The validity of the latter has been demonstrated  by the fits of $\rho(T)/\rho(0)$ to Equ.1 indicated in Fig.2. Then a plot of $\Delta\rho(T)/\rho(T)$ \textit{vs.} $\rho(0)/\rho(T)$ allows to separate the thermal contribution to the AMR, $(\Delta\rho/\rho)_T$, from the defect one, $(\Delta\rho/\rho)_d$. In fact, the extrapolation of the 'high-temperature' data, shown by the inset to Fig.6a, is consistent with a common intercept at $(\Delta\rho/\rho)_T$ = -0.40~$\%$. Such negative contribution to the AMR has been realized early on crystalline Permalloy\cite{BF68} and, more recently, also on polycrystalline Co-films\cite{FGM90}. It was associated with phonon-scattering rather than with magnon contributions. At lower temperatures, our data break away  from the straight lines, which in the thickest film occurs at a rather high temperature, where $\rho(0)/\rho(T)\approx 0.6$. This feature indicates a change of the dominant defect type for scattering  and has also been observed by Freitas \textit{et al.}\cite{FGM90} on various Co-films with different $\rho(0)'s$, i.e. different degrees of disorder.

\subsection{Geometrical Size Effect}

As a guide for discussing the GSE, we refer to Potter's work\cite{Pot74}, who calculated the AMR's produced by majority and minority spins for polycrystalline cubic ferromagnets. He assumed an isotropic scattering potential, as it may be present in the grains of our films. Calculating the sd-scattering rates, Potter considered the effect of the spin-orbit interaction on localized 3d-states, but ignored the influence on the band-structure. Therefore, we expect only a more or less qualitatively correct guidance by infering the AMR-ratio from Ref.\onlinecite{Pot74}:

\begin{equation}
\frac{{\Delta\rho}}{{\rho}} = \frac{{1+r}}{{2+r}}\left\{{\frac{{3\sqrt 3}}{{64}}\left({\frac{{K_{SO}}}{{\in _d}}} \right)^2 - \frac{r}{{560}}\left({\frac{{K_{SO}}}{{2\in_{ex}}}}\right)^2}\right\}    .
\end{equation}
\noindent
Here $K_{SO}\approx ~$0.1~eV measures the spin-orbit coupling energy ${\cal H}_{SO}=K_{SO}~\vec{L}\cdot\vec{S}$. The positive contribution to Equ.5 arises from the longitudinal part of  ${\cal H}_{SO}$ mixing two 3d-orbitals of the minority bands separated by $\in_d$ near the Fermi-surface $\in_F$ . The negative term is due to the nondiagonal part of $\cal{H}$$_{SO}$, which admixes some of the exchange-split majority states to the minority band. The parameter $r=\tau^{-1}_{sd}/\tau^{-1}_{ss}$ accounts for the different scattering rates of the conduction electrons into the 4s- and 3d-states and is mainly determined by the density of states of the 3d-bands at $\in_F$. Because the exchange splitting $2\in_{ex}$ is significantly larger than $\in_d$, the negative majority spin contribution to the AMR may be small relative to the positive one. Taking $r\approx10$ from a recent experiment on Co-films\cite{GSN93},  $\Delta\rho/\rho=(3\sqrt3/64)(K_{SO}/\in_d)^{2}$ follows from Equ.5. Comparing this estimate with our result for the transverse AMR at low temperatures, $\Delta\rho_t(0)/\rho_b(0)\cong10^{-2}$, we obtain for the effective splitting of the two unperturbed 3d-levels, $\in_d\approx3.0~K_{SO}\approx 0.3~\mbox{eV}$. This finding for $\in_d$ becomes smaller if a finite contribution by the majority spins would be considered in Equ.5, but it seems to be reasonable regarding the other simplifying assumptions of the theory\cite{Pot74}. Here we mention the neglect of the effects of the lattice potential and the spin-orbit interaction on the Fermi-surface and on the density of states at $\in_F$\cite{Pot74}, and also of possible hybridizations between the s- and d-orbitals\cite{FC76}.

Nevertheless, we will extend Potter's results derived for an 'isotropic', i.e. polycrystalline cubic ferromagnet, to films with polar texture. Let us recall that the AMR originates from a symmetry breaking of the 3d-orbitals by the magnetization $\vec{M}$ via the spin-orbit coupling. The resulting anisotropic charge distribution gives rise to the scattering asymmetry of the conduction electrons into these 3d-states. On general grounds one may expect that a reduction of the symmetry of the ferromagnet structure weakens the AMR\cite{MP75}, loosely spoken, because then the magnetization induced axial anisotropy of the orbitals becomes less effective. Since the texture in the permalloy\cite{RLC97} and in our cobalt films, both perpendicular to the plane, appear to be strongly correlated with the GSE, we assume the mixing parameter in Equ.5, $k^{2}_{\alpha} =3\sqrt{3}(K_{SO}/4\in_{\alpha})^{2}$, to be different for the in-plane $(\alpha=i)$ and the polar $(\alpha=p)$ directions of $\vec{M}$. Then Equ.5 remains still valid for the in-plane orientations of $\vec{M}$ and ignoring again the small contribution by the majority spins, we have
\begin{equation}
\frac{\rho_\ell - \rho_t}{\rho} =  \frac{1 + r}{2 + r}\quad \frac{3\sqrt{3}}{4}~ k_i^{2}       .
\end{equation}
In order to determine the effect of the film anisotropy on the polar MR, we introduce $k^{2}_{\alpha}$ directly into Potters\cite{Pot74} result for the perpendicular conductivity of the minority spins, $\sigma^{\alpha}_{\bot}/\sigma_0=(3\sqrt{3}/2r)\cdot\ln\left[r/(1+\frac{1}{2}~rk^{2}_{\alpha})\right]$. For small spin-orbit perturbations, $rk^{2}_{\alpha}\ll 1$, the difference between the transverse and polar resistivities becomes:

\begin{equation}
\frac{\rho_t - \rho_p}{\rho} = \frac{1 + r}{2 + r}\quad \frac{3\sqrt{3}}{4} ~ \left( k_i^{2} - k_p{^2}\right)   .
\end{equation}

\noindent
By some trivial algebra we obtain for the GSE from Equs.6, 7:

\begin{equation}
\frac{{\Delta\rho_p }}{{\Delta\rho_t }} = 2 - \frac{{k_p^{2}}}{{k_i^{2}}}  .\\
\end{equation} 

\noindent
Hence, this simple model can explain the upper limit of two of the GSE, which emerges from our data in Fig.6b and also from Fig.6 in Ref.\onlinecite{RLC97} for Permalloy films. Moreover, this model ascribes the GSE to the electronic structure, as it was suspected by Chen and Marsocci\cite{CM72}. Consequently, the GSE should not depend on the temperature which is fully consistent with our results depicted in Fig.6b. 

Equation 8 also predicts that the upper limit of two is reached, when the mixing effect due to the polar oriented magnetization is small compared to mixing by the in-plane $\vec{M}$, i.e. $k^{2}_p<<k^{2}_t$. This case seems to be realized in our films, see Fig.6b, and also in the Permalloy films with increased $\langle 111 \rangle$-epitaxy (Fig.6 of Ref.\onlinecite{RLC97}). These observations indicate that the spin-orbit induced anisotropy in the 3d-orbitals near $\in_F$ is smallest, if $\vec{M}$ is aligned parallel to the existing axial perturbation built in by hcp- or $\langle 111 \rangle$-epitaxy. In this case, the 3d-orbitals have already the axial symmetry so that an induced magnetization along the epitaxial (polar) direction may have only a moderate effect on the scattering probability into these states. This is in distinct contrast to the in-plane orientation of $\vec{M}$ which breaks the symmetry of these orbitals. Therefore, the mechanism proposed here for the GSE qualifies the film anisotropy of the AMR more precisely as structural, rather than as a geometrical effect.

\section{Low-field magnetoresistance}
\subsection{Domain Structures}
\begin{figure}[t]
\center
\psfig{file=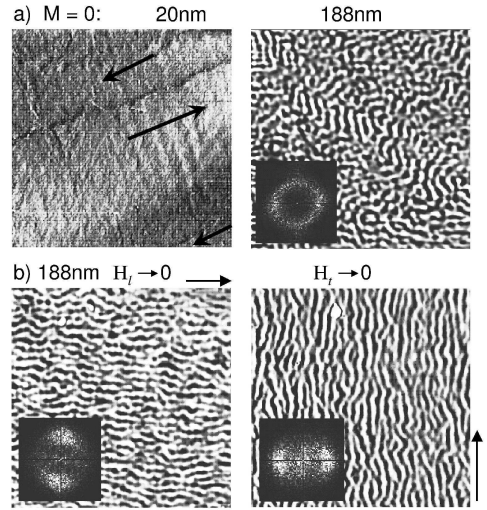, width=7.1cm}
\caption{\protect{Stray field images obtained by magnetic force microscopy: a) In the demagnetized states of 188~nm Co on glass $(10\times10\mu m^{2})$ and 20~nm Co on Si(100) $(25\times25\mu m^{2})$. For the 20~nm film, only domains with in-plane magnetizations along the uniaxial anisotropy field $\vec{H}_u$ are observed, while the 188~nm film displays a maze-structure with an out-of-plane component of ${\vec{M}}$. b) In the remanent states, the 188~nm film reveals stripe domains parallel to the previously applied fields, ${\vec{H}}_{\ell}$ and ${\vec{H}}_t$. The insets show the Fourier-transforms which show the mean size and the directions of the stripe domains.}}
\label{Fig7}
\end{figure}

The formation of domains affects the MR's of the 188~nm thick film and of the thinner films, d$\leq$~20~nm, in quite different ways. The interesting features can already be realized on the large field scale of Fig.5: (i) for d=20~nm Co (and also for 10~nm, not shown) both, the polar and the transverse MR's approach the field-independent longitudinal MR, $\rho_{\ell}$, whereas the polar and the longitudinal MR's of the 188~nm film tend to the field-independent transverse resistance $\rho_t$. In order to provide some solid basis for a detailed discussion of these characteristic features of the domain MR's, we examine the domain structures by magnetic force microscopy (MFM).

The essential difference between the thick (188~nm) and the thinner films can be infered from MFM images of the demagnetized states, shown in Fig.7a. The images have been recorded in the dynamic mode of the Q-scope which is sensitive to the polar gradient of the polar force, i.e. $\delta F_p/\delta x_p=M_p~\delta^{2}H_p/\delta x^{2}_p$. The 20~nm film consists of large, some 10~$\mu m$ wide domains with in-plane magnetizations separated by 180$^{\circ}$ Neel walls. The domain magnetizations are oriented parallel to the uniaxial anisotropy field $\vec{H}_u$ as determined by FMR. A slight longitudinal ripple of $\vec{M}$ about $\vec{H}_u$ is visible, which most likely arises from the polycrystallinity of the film. 

\begin{figure}[t]
\center
\psfig{file=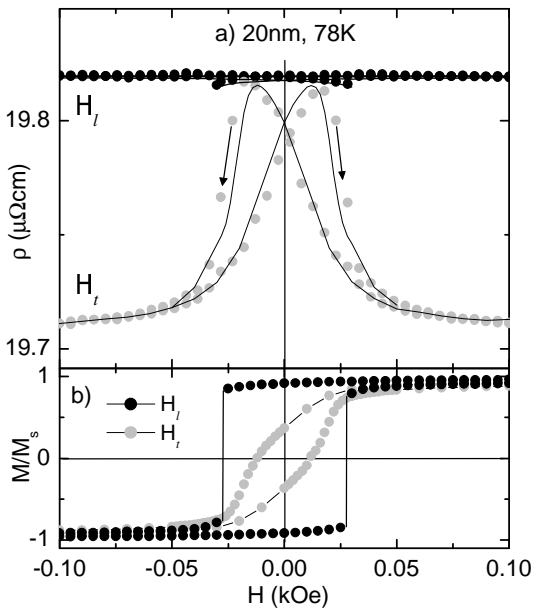, width=7.1cm}
\caption{\protect{a) Low-field MR of 20~nm Co measured at 78~K for in-plane fields applied longitudinally $(H_{\ell})$ and transversely $(H_t)$ to the current. The solid curves are fits to the AMR effect (Equs.2,9) using the magnetization curves $M(H_{\ell})$ and $M(H_t)$ as shown in panel b). The longitudinal field has been applied along to the growth-induced uniaxial anisotropy field $H_u$~=~20~Oe determined by ferromagnetic resonance\cite{Gil04}.}}
\label{Fig8}
\end{figure}

In constrast, the 188~nm film exhibits a maze configuration of stripe domains with sizable polar components of the domain magnetizations. The Fourier transform of the image in Fig.7a yields a mean width of the stripes, $d_D=(205\pm15)$~nm, being rather close to the thickness as expected for weak stripes by magnetostatic reasons \cite{HS98}. Recently, the same observations were reported for a 195~nm thin \textit{polycrystalline} Co-film on glass and related to a hexagonal texture perpendicular to film plane\cite{KCB04}. MFM images depicted on \textit{epitaxial} Co-films  revealed a reorientation of the domain magnetization from in-plane to polar between 10~nm and 50~nm ~\cite{HPO96, DPH02} which was explained in terms of the perpendicular magnetocrystalline anisotropy of Co. These results suggest that also in our case the hcp texture, realized by the XRD (Fig.1), generates such a crystalline anisotropy, which in the 188~nm film becomes large enough to produce a significant polar component of $\vec{M}$. Let us also recall that we supposed this texture already in the discussion of the GSE.

The other interesting property of these weak stripes is seen in Fig.7b. In the remanent states, stripe patterns are found aligned with the direction of the previously applied fields $H_{\ell}$ or $H_b$. This socalled rotable anisotropy can be attributed to the stiffness of the domain walls against deformations~\cite{HS98} and is probably supported by a pinning of the walls by local anisotropies in the granular structure. The rotatable anisotropy suggests also an 'isotropic' hysteresis loop, the shape of which should be independent on the direction of the in-plane field. In fact, we do observe this feature on the 188~nm film, see Fig.9 below, and will refer to it when discussing the MR in the domain state.

\subsection{Anisotropic Magnetoresistance}

We begin with the low-field resistance of the \textit {thin} films, exemplified by Figs.3a, 5a for d=20~nm: both the transverse and the polar MR's, $\rho(H_t\rightarrow 0)$ and $\rho(H_p\rightarrow 0)$, tend to the longitudinal one, $\rho(H_{\ell})$. This behavior is readily explained by the fact that the resistance is measured along $\vec{H}_u$, and that at low fields the domain magnetization is also directed parallel to $H_u$ evidenced by MFM (Fig.7a). The parabolic decrease of $\rho$ in larger \textit {polar} fields, $\Delta\rho(H_p<M_s)\sim -H^{2}_p$, was already attributed to the AMR resulting from the rotation of $\vec{M}$ from an in-plane to the polar direction. 
Also the detailed variation of the \textit {in-plane} MR's, shown in Fig.8a, can be explained by the AMR. Using the hysteresis loops $M(H_i)$ in Fig.8b, and assuming the relations for the angle $\varphi$ in Equ.2,
\begin{subequations}
\begin{equation}
\cos \varphi \left( {H_\ell  } \right) = M\left( {H_\ell  } \right)/M_s,
\end{equation}
\begin{equation}
\sin \varphi \left( {H_t } \right) = M\left( {H_t } \right)/M_s ,
\end{equation}
\end{subequations}

\noindent
the in-plane MR can be described rather nicely. The physical arguments for these agreements are: (i) the {\itshape longitudinal} magnetization process, $M(H_{\ell})$, is due to the nucleation of 180$^{\circ}$ Neel walls (see Fig.7a) at the coercive field $H_c= -H_u$ (determined by FMR), which then rapidly cross the film leaving the resistance unchanged; (ii) upon reduction of the {\itshape transverse} field, on the other hand, a longitudinal ripple of $\vec{M}$ about $\vec{H}_t$ appears which originates from $\vec{H}_u$ (see e.g. Ch.5.5 of Ref.\onlinecite{HS98}). Accordingly, the components of $\vec{M}$ parallel and antiparallel to the current $\vec{I}||\vec{H}_u$ are growing continuously so that $\rho(H_t)$ increases until the transverse coercive field $H^{t}_c<H_u$ is reached. There the magnetization component along $\vec{H}_u$ changes sign and increases at the expense of the ripple, so that $\rho(H_t)$ back to $\rho_t$ at larger negative fields.

\begin{figure}[t]
\center
\psfig{file=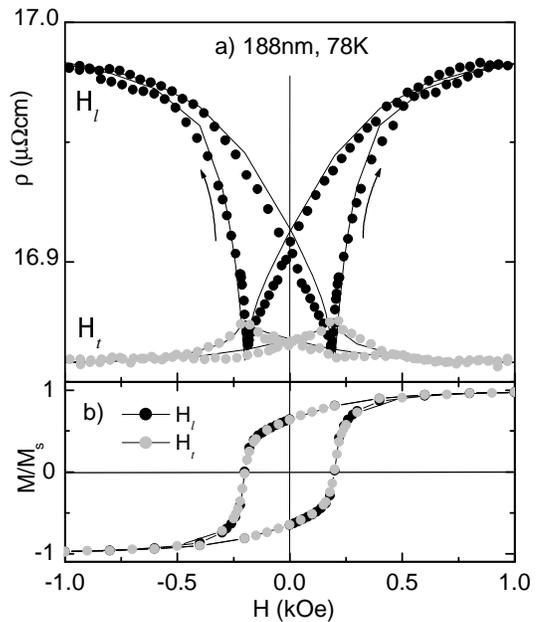, width=7.1cm}
\caption{\protect{a) Low-field MR of 188~nm at 78~K in longitudinal and transverse fields. As in Fig.7, the solid curves are fits to Equs.2,9 using the magnetization curves displayed in panel b). Note the inversion of the longitudinal and transverse field MR-variations in comparison to the 20~nm film, shown in Fig.8a.}}
\label{Fig9}
\end{figure}

A rather different behavior is displayed by the 188~nm \textit {thick} film. Already in Fig.5b we noticed that at low fields the polar and the longitudinal MR's tended to the transverse MR. As a rather unexpected feature, the transverse MR turned out to be nearly independent of the field also in the domain regime, $\rho(H_t)=\rho_t$. The detailed variation of the in-plane MR's at low fields is shown in Fig.9a revealing just the opposite to the behavior of the thin films (see Fig.8a): the longitudinal MR displays a strong field dependence, while the transverse MR remains very small. These results are explained also by the AMR effect. The \textit{in-plane} MR, shown in Fig.9b, is rather nicely reproduced by the solid curves which have also been calculated from Equ.2. Again, the mean angle $\varphi$ between current and magnetizations $\vec{M}(\vec{H})$ has been determined from Equ.9 and the hysteresis loops, Fig.9b. As a matter of fact, we emphasize that the shape of these loops does not depend on the direction of the in-plane field ('rotatable loops'). This is consistent with the corresponding behavior of the weak stripe domains depicted by MFM in Fig.7b. In contrast to the thin films, d$\leq$20~nm, no effect by the uniaxial in-plane anisotropy field, $H_u=15$~Oe, determined by FMR~\cite{Gil04}, is realized. The much larger coercive field, $H_c\approx 200$~Oe, stems most likely from the pinning of the stripe domain walls by the random polycrystalline anisotropy in the films.

\subsection{Effective Medium Approach}

Aiming at a more detailed description of the MR in the 188~nm film, again the domain structure has to be taken into account. For this purpose, we use an effective medium model, by which R\"udiger \textit{et al.}\cite{RYT99} successfully interpreted the AMR of epitaxial Co-films. Introducing the volume fractions $v_i$ for different domain species, the AMR is approximated by 

\begin{equation}
\Delta \rho (\vec H) = \sum\limits_{i = 1}^3 {} \upsilon _i (\vec H)\,\Delta \rho _i .
\end{equation}

\noindent
Here the $\Delta\rho_i$ denote the AMR's of the corresponding domain with polar, transverse or longitudinal orientations of $\vec{M}$ relative to current and film plane. By definition is $\Delta\rho_{\ell}=0$ and if, for convenience, $\Delta\rho(\vec{H})$ is normalize to the transverse MR, Equ.10 takes the form

\begin{equation}
\frac{\Delta \rho ( \vec H )}
{\Delta \rho _t } = \upsilon _t( \vec H ) + g_s \upsilon _p ( \vec H ),
\end{equation}

\noindent
where $g_s=\Delta\rho_p/\Delta\rho_t$ denotes the GSE-ratio. The simplest case, $\upsilon_t=\upsilon_p=0$ and, hence, $\Delta\rho=0$ has been realized on the thin films at low fields. 

The most interesting example is the 188~nm film, where (i) the low-field MR appears to be inverted relative to the thin films and, moreover, (ii) the transverse resistivity remains at the saturation value $\Delta\rho(H_t)=\Delta\rho_t$, even in the domain state. We will now attempt to relate these striking features displayed by Figs.5b, 9 to the domain structure observed by MFM, see Fig.7. Observation (ii) in connection with Equ.11 implies for the concentration of polar oriented domains,
\begin{equation}
\upsilon _p(H_t) = \frac{1}{g_s }\,\left[ 1 - \upsilon _t(H_t) \right] .
\end{equation}
\noindent
Below the saturation field, the magnetization $M(H_t)$ and therefore, $\upsilon_t(H_t)$, starts to decrease at the expense of a finite polar component $\upsilon _p$, which leads to the nucleation of stripe domains. Upon further reduction, $H_t\rightarrow 0$, the hysteresis loops display a normalized remanent magnetization $M(H_t\rightarrow 0)/M_s=0.66(2)$, i.e. volume fraction $\upsilon_t(0)\cong 2/3$. For an estimate, we take the maximum GSE, $g_s=2$, to find from Equ.11 $\upsilon_p(0)=1/6$ and by using $\sum\limits_{i = 1}^3 {\upsilon _i  = 1}$,the same longitudinal volume fraction $\upsilon _\ell  (0) = 1/6 = \upsilon _p (0)$. The agreement of both volumes implies that the nucleation of \textit{polar} domains is accompanied by the creation of an equal amount of \textit{longitudinally} oriented domain. Considering the square-like cross-section of the stripes following from Fig.7b, this result indicates that the flux extending from the polar phase is closed by the longitudinal volume $\upsilon_{\ell}(-H_c)$. The rotatable symmetry of the hysteresis loops implies for the longitudinal direction also $\upsilon_p(H_{\ell}\rightarrow 0)=\upsilon_t(H_{\ell}\rightarrow 0)=1/6$. For the longitudinal MR Equ.10 predicts then $\Delta\rho(H_{\ell}\rightarrow 0)/\Delta\rho_t=1/6+2\cdot1/6=1/2$, which is in close agreement with the measured value, see Fig.9a.

Finally, upon reduction of $H_t$ to the coercive field, $\upsilon_p(H_t)$ increases further. The volume fraction of the polar domains at $-H_c$ can be estimated from  the stripe maze of the demagnetized state, Fig.7a, which suggests $\upsilon_t(-H_c)=\upsilon_{\ell}(-H_c)$. Then, from Equ.11 and simple algebra we obtain $\upsilon_p(-H_c)=1/3=\upsilon_{\ell}(-H_c)=\upsilon_{\ell}(-H_c)$. Thus the demagnetized state consist of equal volumes for all six possible magnetization directions, which by considering the symmetry of the stripe structure is again a plausible result.

\section{Summary and Conclusions}
The magnetoresistance of polycrystalline Co-films, which were characterized by XRD, FMR, SQUID-magnetometry, AFM, MFM and temperature variable resistivity, has been investigated in fields up to 100~kOe directed along three principal directions. In the saturated state, the MR displayed the socalled geometrical size-effect (GSE), according to which the MR for the polar orientation of $\vec{M}$ is up to twice as large as for the in-plane $\vec{M}$. The determination of the GSE was facilitated by the facts that the spin-wave contributions could be easily subtracted and that in the present disordered films the classical Lorentz MR proved to be negligible. Basing on a correlation between the GSE and a texture detected previously on Permalloy films\cite{RLC97} and also on our Co-films by XRD, we proposed to attribute the GSE to an anisotropic mixing of the 3d-levels near $\in_F$ by the longitudinal part of the spin-orbit interaction. By extending Potter's\cite{Pot74} prediction for the AMR of the minority spin channel, we obtained a result which is consistent with the observed upper limit of two for the GSE and also with the temperature independence of the GSE. A relation of the GSE to the electronic structure has already been conjectured in the literature\cite{CM72}, but not yet been worked out. Of course, regarding the simplicity of the proposed extension and the assumption of a simple spherical Fermi-surface for the final 3d-states in Ref.\onlinecite{Pot74}, which considers only the local aspect of the spin-orbit interaction, these consistencies may be fortuitous. However, we believe that the central argument for the appearance of the GSE, i.e. the presence of an additional uniaxial symmetry in polycrystalline films through a texture in thin films remains valid. Hence, a more detailed reasoning for the GSE considering also the effect of the spin-orbit interaction on the band-structure is indicated in order to check the present rough model.

The MR's in the \textit{domain state} were interpreted using the saturated AMR's and the GSE, the hysteresis loops, and MFM images of the domain structure. For thin Co-films, $d\leq20~\mbox{nm}$, where the magnetization remained the film-plane and became for $\vec{H}\rightarrow0$ parallel to the weak uniaxial anisotropy field, the MR attained the maximum (longitudinal) value at zero magnetization. The MR in the domain state of the thickest film, d=188~nm, on the other hand, displayed a rather different behavior. As a function of the \textit{transverse} field in the film-plane, the resistance turned out to be almost constant, whereas upon reducing \textit{longitudinal} and \textit{polar} fields the resistances decreased and increased, respectively, from their different saturation values to the transverse MR. MFM images and hysteresis loops revealed the formation of rotatable stripe domains with square cross-section due to the hcp texture. By means of an effective medium model\cite{RYT99}, the MR's could be quantitatively explained in terms of a flux closure configuration of the magnetization components about the directions of the stripes. Approaching the coercive field, the stripes terminated in a maze configuration, and the fractional volumes of all three magnetization components proved to be equal. In this model, the surprising field independence of the transverse MR results from the squared corss-section of the stripes with transverse flux-closure and from a GSE ratio of two. We should note that this discussion did not invoke (possible small) contributions to the MR by the Neel- and Bloch-walls in the thin and thick films, respectively. Such effects have been reported before in epitaxial Co-films\cite{GAO96,VVC96,RYT99} with strong hcp crystalline anisotropy and quantitatively different domain dimensions.

The authors are indebted to the late Prof. J. Appel (Hamburg) for his encouraging interest during the early steps of this work. Discussions with Proff. P. B\"oni (M\"unchen), D. Grundler (Hamburg) and M. Morgenstern (Aachen) are gratefully acknowledged.


\end{document}